\documentclass[12pt]{iopart}
\usepackage[numbers,sort&compress]{natbib}
\usepackage{lineno}
\usepackage{graphicx}
\usepackage[labelfont=bf,font=scriptsize]{caption}
\usepackage{hyperref}
\usepackage{booktabs}  
\usepackage{changepage,threeparttable} 
\usepackage{enumitem}

\usepackage{iopams}
\usepackage{siunitx}
\makeatletter
\long\def\@makefntext#1{\parindent 1em\noindent 
 \makebox[1em][l]{\footnotesize\rm$\m@th{^\arabic{footnote}}$}%
 \footnotesize\rm #1}
\def\@makefnmark{\hbox{${^\arabic{footnote}}\m@th$}}
\def\@thefnmark{\arabic{footnote}}
\makeatother

\begin{document}
\title[Dark Solitons in Bose-Einstein Condensates Dataset]{Dark Solitons in Bose-Einstein Condensates:\\
A Dataset for Many-body Physics Research}

\author{Amilson R. Fritsch\textsuperscript{1}, Shangjie Guo\textsuperscript{1}, Sophia M. Koh\textsuperscript{2}, I.~B.~Spielman\textsuperscript{1,3}, Justyna P. Zwolak\textsuperscript{3,*}}
\address{
\textsuperscript{1}Joint Quantum Institute, National Institute of Standards and Technology, and University of Maryland, Gaithersburg, MD 20899, USA}
\address{\textsuperscript{2}Department of Physics and Astronomy, Amherst College, Amherst, MA 01002, USA}
\address{\textsuperscript{3}National Institute of Standards and Technology, Gaithersburg, MD 20899, USA}
\ead{jpzwolak@nist.gov}

\begin{abstract}
We establish a dataset of over $1.6\times10^4$ experimental images of Bose--Einstein condensates containing solitonic excitations to enable machine learning (ML) for many-body physics research.
About $33~\%$ of this dataset has manually assigned and carefully curated labels. 
The remainder is automatically labeled using SolDet---an implementation of a physics-informed ML data analysis framework---consisting of a convolutional-neural-network-based classifier and OD as well as a statistically motivated physics-informed classifier and a quality metric.
This technical note constitutes the definitive reference of the dataset, providing an opportunity for the data science community to develop more sophisticated analysis tools, to further understand nonlinear many-body physics, and even advance cold atom experiments.
\end{abstract}
\vspace{2pc}
\section{Introduction}\label{sec:intro}

Advances in machine learning (ML), and especially in the area of deep learning, are data driven.
Yet, in many fields of science it is a common practice for researchers to either make data available only upon ``reasonable request,'' or to not share it at all. 
As a result, the development of specialized ML techniques as well as trained models is often stymied by the lack of high-quality relevant datasets.
Cold atom experiments produce vast amounts of data---images of atom clouds---making cold atoms an ideal system where data availability can both advance ML research and provide new applications relevant to experiment~\cite{PhysRevApplied.14.014011,Seo21-MAN,Leykam22-DPH,Guo22-CMP}.
Here we provide a two component dataset consisting of absorption images of dark solitons in atomic Bose--Einstein condensates (BECs).
The first component of this dataset is a curated revision of the {\it Dark solitons in BECs dataset v.1.0}~\cite{solitons-data,Guo21-MDS} containing over $6\times10^3$ images, which we carefully amend to assure high quality labels.
The second component contains approximately $1\times10^4$ additional preprocessed and automatically labeled images.

BECs are widely investigated systems that exhibit quantum phenomena on a macroscopic scale.
For example, they can be manipulated to contain solitonic excitations including conventional solitons, vortices, and many more.
Broadly speaking, solitonic excitations are solitary waves that retain their size and shape and often propagate at constant speed.
They are present in many systems, at scales ranging from microscopic~\cite{Burger1999, Denschlag97}, to terrestrial~\cite{Russel1837,Hasegawa1973,Osborne1980, PhysRevLett.45.1095,Yomosa1984, Hashizume1985, Lakshmanan2009Tsunamis} and even astronomical~\cite{Stasiewicz2003}.

Unlike naturally occurring physical systems, the parameters governing BECs are under strict experimental control.
In atomic BECs, solitons can be classified as dark or bright, generally occurring for systems with repulsive or attractive interactions, respectively. 
Here, we focus on repulsively interacting BECs that therefore support dark solitons. 
In one-dimensional (1D) BECs, only a single type of dark soliton exists, corresponding to the kink soliton in three-dimensional (3D) systems.
Even in highly elongated 3D BECs, kink solitons are stable only when strict conditions on the propagation velocity and the trap geometry are satisfied.
Otherwise they decay, ultimately producing solitonic vortices, stable solitonic excitations in 3D BECs.

Kink solitons manifest as a plane-like reduction in the BEC's 3D density, while solitonic vortices add to this reduction a vortex-line of zero density about which the BEC flows (with either sense of vorticity).
Visually, kink solitons and solitonic vortices both appear as depletions in the BEC's atomic density, with solitonic vortices having additional structure that is not present for kink solitons.
Given this, one might expect such images of BECs to be straightforward to classify.
However, due to physical limitations of the imaging system capturing only a single-sided view of the 3D BEC, they are not.
For example, the plane of reduced density can be canted with respect to the BEC, both for kink solitons and solitonic vortices (leading to our label ``canted excitation'').
The solitonic vortex~\cite{Tylutki15-SVC,Donadello2014solitvort} can be predominantly confined to the top or the bottom of the BEC (leading to what we label as ``top partial'' and ``bottom partial'' excitations, respectively); in addition, vorticity leaves its imprint on density as well (giving our labels ``clockwise vortex''  and ``counterclockwise vortex'').
Finally, when viewed from the side, the density-zero of a solitonic vortex becomes indistinguishable from a shallow kink soliton~\cite{Donadello2014solitvort} (leading to our joint label of ``longitudinal soliton'').

The expanded dataset, {\it Dark solitons in BECs dataset v.2.0}, contains experimental images of dark solitons created using the procedure described in reference~\cite{Fritsch20-SCV}.
The constituent images reflect a diverse range of parameters, for example containing from zero to upwards of four excitations of numerous types that themselves have a wide range of velocities.
Due to the similarities between kink solitons and other excitations present in our images, annotators can potentially mislabel data, attributing different labels to the same type of excitation. 

The full dataset contains over $1.6\times10^4$ experimental images, including $5\,378$ images manually classified into three carefully curated classes: ``no excitations'' (class-0), ``lone excitation'' (class-1), and ``other excitations'' (class-2).
The lone excitation class is additionally tagged with the excitation position, the physics-informed excitation class (PIE class), and quality score. 
The remaining images are not manually classified; rather they are automatically labeled using the SolDet package~\cite{SolDet}. 
This dataset is available via the National Institute of Standards and Technology Public Data Repository~\cite{solitons-data} to provide an opportunity for the data science community to develop more sophisticated analysis tools for soliton research and to further understand nonlinear many-body physics.

\section{Dataset curating: materials and methods}\label{sec:methods}
In 2021, we released the ``Dark solitons in BECs dataset'' \cite{solitons-data} consisting of approximately $6.3\times10^3$ preprocessed absorption images taken from multiple experiments performed in a single lab over a span of two months with human assigned labels.
Based upon the number of solitonic excitations observed in a given image of a BEC, the data was organized into three classes: no excitations (class-0, accounting for $19.8~\%$ of the dataset), lone excitation (class-1, accounting for $55.4~\%$), and other excitations (class-2, accounting for $24.8~\%$).
While the initial agreement rate between annotators was relatively high at $87~\%$, the remaining $13~\%$ of the dataset had to be ``further analyzed and discussed until an agreement'' between annotators was reached\footnote[4]{The agreement was substantially higher for the easier to interpret class-0 ($95.7~\%$) than for other two cases ($88.7~\%$ for class-1 and $76.3~\%$ for class-2), indicating a likely decrease in label reliability with increased data complexity.}, as stated in reference~\cite{Guo21-MDS}.
Such discussions of labels might introduce an undesirable bias in the labels, especially when the data is challenging to interpret.
This bias is in turn imprinted into any ML model trained using that data, thereby putting the model's reliability into question. 

While manually re-examining the dataset in the context of reference~\cite{Guo22-CMP}, we confirmed the presence of inconsistencies in the human assigned labels.
We found three types of labeling errors:
\begin{enumerate}
    \item[(e1)] Images in class-2 containing only a single excitation.
    \item[(e2)] Images in class-1 containing more than one solitonic excitation.
    \item[(e3)] Images in class-0 containing a distinct excitation.
\end{enumerate}  
Moreover, in reference~\cite{Guo21-MDS}, the excitation's location for the lone excitation class was determined using fits centered on the deepest density depletion.
Our reexamination showed that some solitonic excitations were far from this point [error type (e4)]. 
As a result, we decided to use a combination of ML and statistical analyses to identify potentially incorrectly labeled data as well as to curate the dataset.

Building on the deep ensembles approach, originally proposed as a means to estimate models' predictive uncertainty~\cite{Lakshminarayanan17-UDE}, we implement an iterative five-fold cross-validations\footnote{The $k$-fold cross-validation is a resampling method that involves dividing the full dataset into $k$ partitions and then performing a series of training and testing runs, with each run using $k-1$ partitions to train a given module and the remaining one partition to test it. 
The process is repeated $k$ times to fully cover the dataset.} with strict agreement constraints to curate the original dataset.
As described in detail below, we employ two classifications schemes for the cross-validation.
First, we verify the human assigned labels using a set of convolutional neural network (CNN) classifiers trained using the original dataset.
In addition, to ensure diversity of the trained models~\cite{Ovadia19-CYU}, we use a set of object detectors (ODs)~\cite{Redmon16-LOO,voulodimos2018deep} trained to localize all solitonic excitations within each BEC image and compare the number of detected excitations with both the original and CNN labels.
After each iteration, images with insufficient agreement are further analyzed and, if necessary, removed from the dataset and set aside. 
Following the data curation process, we add the location of all excitations obtained from the ODs as an additional label.

The curated dataset is used to train an implementation of SolDet~\cite{SolDet}, a general-purpose framework for feature identification in cold atom experiments~\cite{Guo22-CMP}.
We then use the PIE classifier SolDet module to add fine-grained and physically-motivated labels (e.g., longitudinal solitons, solitonic vortices, and  ``partial'' solitons) for the lone excitation class.
Furthermore, we employ SolDet to automatically label about $1\times10^4$ additional experimental images (class-9) as well as all images set aside during the curation process (class-8)~\cite{solitons-data}.

\begin{figure}[t]
    \centering
    \includegraphics[width=0.99\linewidth]{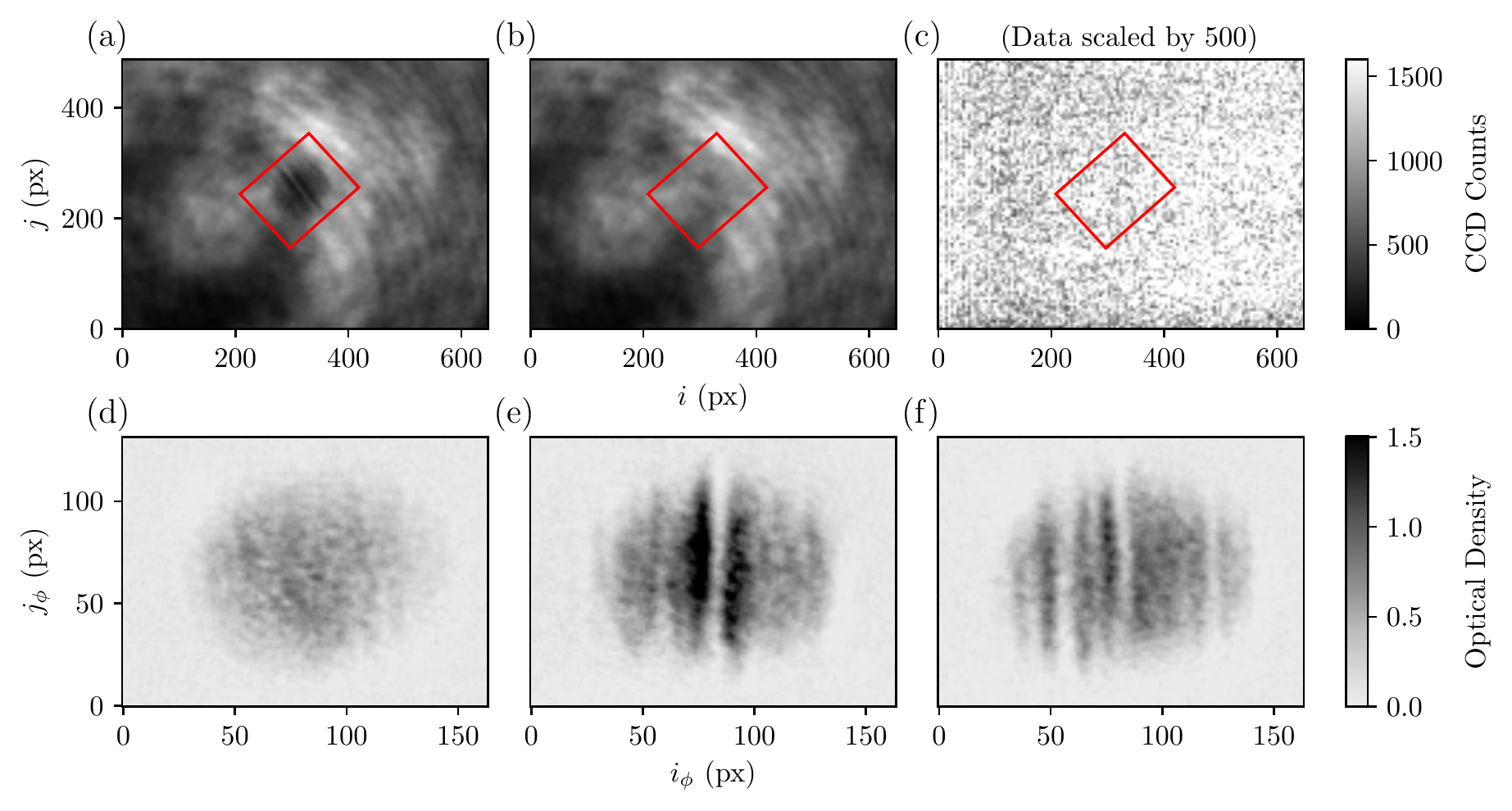}
    \caption{
    (a)--(c) Raw data from which dataset element \texttt{2019-07-12\_0061\_20190523\_BEC\_F1\_NewODT\_DMD\_365} (shown in (e)) is derived: (a) with atoms $I^{\rm A}$, (b) probe only $I^{\rm P}$, and (c) background $I^{\rm BG}$ (data scaled by 500). 
    The red boxes in each raw image indicate the region where the BEC is located.
    (d)--(f) Representative preprocessed absorption images from the curated dataset. 
    (d) No soliton [element: \texttt{2019-07-15\_0058\_20190523\_BEC\_F1\_NewODT\_DMD\_331}], 
    (e) single soliton, and 
    (f) other excitations [element: \texttt{2019-09-25\_0080\_20190910\_BEC\_F1\_NewODT\_DMD\_10}].}
    \label{fig:soliton-sample}
\end{figure}

\subsection{Data preprocessing}\label{ssec:dat-proc}
In the raw data, shown in figures~\ref{fig:soliton-sample}(a)--(c), the BEC occupies only a small region of the image (inside the red box) and the long axis of the BEC is rotated with respect to the camera.
The horizontal and vertical axes in figure~\ref{fig:soliton-sample}(a-c) are labeled in terms of camera pixels $i$ and $j$.

The angle between the camera and the BEC depends on the experimental setup and is obtained from fits to a representative subset of absorption images (in the case of our data the rotation angle is about $\phi=40$ degrees).
The BEC's angle, position, and size---all necessary for proper cropping---are determined by fitting every image to a column-integrated 3D Thomas-Fermi distribution 
\begin{equation}\label{eq:2Dfit}
n^{\rm TF}_{i,j} = n_{0}\,{\max}\left\{\left[1-\left(\frac{i_\phi}{R_i}\right)^2-\left(\frac{j_\phi}{R_j}\right)^2\right],0\right\}^{3/2} + \delta n,
\end{equation}
describing the density distribution of 3D BECs integrated along the imaging axis~\cite{Castin1996}.
There are seven parameters in this fit: the rotation angle $\phi$; the BEC center coordinates $[i_0, j_0]$ in the original image frame; the peak 2D density $n_0$; the Thomas--Fermi radii [$R_i$, $R_j$]; and an offset $\delta n$ (from small changes in probe intensity between images).
We define the rotated coordinates as $i_{\phi} = (i-i_0) \cos(\phi) + (j-j_0) \sin(\phi)$ and $j_{\phi} = (j-j_0) \cos(\phi) - (i-i_0) \sin(\phi)$.
The initial estimates for $[i_0, j_0]$ are obtained by summing the 2D density image along the vertical and horizontal directions to obtain two 1D projections and selecting the average position of the five largest values.
The largest value of the image is taken as the estimate for $n_0$ and $[R_i, R_j] = [66,55]$ as the estimate for the radii.
The estimate for the offset $\delta n$ is zero.
The $164\times132$ pixel extent  of the cropping region 
is determined from the average radii $[R_i, R_j] = [66(5), 58(3)]$ obtained from fits to the original $6.2\times10^3$ images\footnote{We use a notation value(uncertainty) to express uncertainties, for example $1.5(6)\ {\rm cm}$ would be interpreted as $(1.5\pm0.6)~{\rm cm}$. All uncertainties herein reflect the uncorrelated combination of single-sigma statistical and systematic uncertainties.}. 
In the resulting images, dark solitons appear as vertically aligned density depletions and are often easily visually identified, see figures~\ref{fig:soliton-sample}(e) and (f).

To ease the manual analysis and labeling process and facilitate training of the ML models, the absorption images are first rotated to align the BEC with the image frame and then cropped to discard the large fraction of the image that does not contain relevant information.
Finally, an elliptical mask (determined based on the $[R_i, R_j]$ radii) is applied to the image to eliminate the noise outside the BEC~\cite{Guo21-MDS}.

The same preprocessing techniques are applied to the over $1\times10^4$ previously unlabeled absorption images now included in the expanded ``Dark solitons in BECs dataset 2.0''.

\subsection{Data labeling process: dark solitons in BECs dataset v.1.0}\label{sec:labeling-v1}
As discussed in reference~\cite{Guo21-MDS}, the {\it Dark solitons in BECs dataset v.1.0}~\cite{solitons-data} consists of images labeled by three independent annotators.
These labels organized the data into three disjoint classes: class-0 indicating no excitations, class-1 indicating lone excitation, and class-2 indicating other excitations (e.g. different types of excitations, multiple excitations, ambiguous data).
The expectation was for class-0 to consist of images that unambiguously contain no solitonic excitations; for the class-1 to contain images with exactly one solitonic excitation; and for class-2 to contain all other images.

The initial labeling process was carried out in batches.
At each stage, a subset of anywhere between 508 and 1209 images were independently labeled by each annotator and the resulting labels were compared.
The labels with full agreement were accepted.
When only two out of three annotators agreed (moderate disagreement), the images were reinspected and further discussed until an agreement was reached.
Finally, images labeled differently by each annotator (strong disagreement), were added to class-2. 
The top two rows in table~\ref{tab:labeling} show the distribution of images between classes in {\it Dark solitons in BECs dataset v.1.0}, as well as the number of images with full agreement between annotators in each class.

\begin{table}[t]\centering
\scriptsize
\captionsetup{width=\linewidth}
\caption{Labeling statistics.  
The first two rows describe the labeling distribution in {\it Dark solitons in BECs dataset v.1.0} and include the number of images with full agreement in each class.
The third row counts the number of images in each class that pass the first round of CNN and OD tests.
The last row shows the number of images remaining in each class in the curated dataset.}
\begin{tabular*}{\textwidth}{l@{\extracolsep{\fill}}cccccc}\toprule
Labeling phase & Class-0 & Class-1 & Class-2 & Class-8 & Class-9\\\midrule
Original dataset & 1\,237 & 3\,468 & 1\,552 & --- & ---  \\
Original dataset: 3-agree & 1\,184 & 3\,077 & 1\,184 & --- & ---\\ 
CNN and OD test & 1\,234 & 3\,449 & 1\,388 & 186 & --- \\ 
Curated dataset & 1\,130 & 3\,212 & 1\,036 & 879 & 10\,221 \\
\bottomrule
\end{tabular*}
\label{tab:labeling}
\end{table}

\subsection{Data labeling process: Dark solitons in BECs dataset v.2.0}\label{sec:labeling-v2}
During the first phase of data curating, we performed a pair of five-fold cross validation tests using the original dataset.
The first cross validation used CNN models trained on all three classes and the second used ODs trained with only class-0 and class-1.
After cross validation we tagged each excitation located by the OD with a quality estimate~\cite{Guo22-CMP}.
The quality estimator yields the likelihood that a fit to the 1D profile of a given excitation has parameters in the range expected for a solitonic excitation.
The likelihood was established based on a statistical analysis of fits to features previously identified as solitonic excitations in comparison with all other density depletions.

The cross-validations results in each image in class-0 and class-1 being assigned two predicted labels that are used to identify ambiguous data.
To enable direct comparison of the two models, the OD predicted label is class-0 if no excitations are identified, class-1 if one excitation is found, and class-2 in all other cases.
An image is flagged as potentially mislabeled if and only if
\begin{enumerate}
    \item the CNN prediction disagrees with the assigned class and 
    \item the CNN and OD predictions are the same.
\end{enumerate}
We found $343$ potentially mislabeled images: $14$ in class-0, $40$ in class-1, and $289$ in class-2.
We note that our intent is to use cross-validation and deep ensembles to assist data curating, not to change the ground truth.
Thus, we do not overwrite the original class labels during the curating process.
Rather, all flagged images are further analyzed and all potential excitations are tested with the quality metric.
At this stage, all images in the extended {\it Dark solitons in BECs dataset v.2.0} have \texttt{label\_v1} (either the original label or class-9 for new, therefore unlabeled, data), and in addition they are assigned a new intermediate label \texttt{label\_v2} (see table~\ref{tab:data-dictionary}).
\texttt{label\_v2} is set equal to \texttt{label\_v1} except for data determined to be truly mislabeled where  \texttt{label\_v2} is assigned to a new class-8 effectively removing it from the curated dataset.
The resulting distributions between classes is shown in table~\ref{tab:labeling}.

\begin{table}[t]\centering
\scriptsize
\captionsetup{width=\linewidth}
\caption{Complete dictionary of the labels appearing in {\it Dark solitons in BECs dataset v.2.0}.
For each element we provide the key, definition together with all possible instances of a given label, and data type.
The \texttt{excitation\_position}, \texttt{excitation\_PIE}, and \texttt{excitation\_quality} labels are assigned only to data where \texttt{label\_v3}=1.
The SolDet labeles (\texttt{soldet\_CNN}, \texttt{soldet\_OD}, \texttt{soldet\_PIE}, and \texttt{soldet\_QE}) are assigned to all of the class-8 and class-9 data as well as to the 10~\% of the curated manually labeled data used for testing during SolDet training. 
}
\begin{tabular*}{\textwidth}{p{0.17\linewidth}p{0.64\linewidth}p{0.12\linewidth}}
\toprule
Dictionary key & Definition & Data type \\\midrule
\texttt{file\_name} & Information about which data file a given set of labels refers to & String \\
\texttt{label\_v1} & The original, human assigned label as in {\it Dark solitons in BECs dataset v.1.0} 
\begin{itemize}[nosep]
    \item[0:] no excitations
    \item[1:] lone excitation
    \item[2:] other excitations 
    \item[9:] unlabeled \vspace{-0.5\baselineskip}\mbox{}
\end{itemize} & Integer \\
\texttt{3-agree} & Indicates whether all annotators agreed on the originally assigned label 
\begin{itemize}[nosep]
    \item[1:] true
    \item[0:] false
    \item[-1:] unlabeled data \vspace{-0.5\baselineskip}\mbox{}
\end{itemize} & Integer  \\
\texttt{label\_v2} & Intermediate label resulting from the first curation phase 
\begin{itemize}[nosep]
    \setlength{\itemindent}{.25in}
    \item[0,1,2,9:] same as for \texttt{label\_v1}
    \item[8:] data determined to be potentially mislabeled \vspace{-0.5\baselineskip}\mbox{}
\end{itemize} & Integer \\
\texttt{label\_v3} & Curated label 
\begin{itemize}[nosep]
    \setlength{\itemindent}{.32in}
    \item[0,1,2,8,9:] same as for \texttt{label\_v2}
    \vspace{-0.5\baselineskip}\mbox{}
\end{itemize}  & Integer \\
\texttt{excitation\_position} & True position of the excitation for the lone excitation class & List \\
\texttt{excitation\_PIE} & Physically-motivated label provided by the PIE classifier
\begin{itemize}[nosep]
    \item[A:] longitudinal soliton (0)
    \item[B:] top partial (1)
    \item[C:] bottom partial soliton (2)
    \item[D:] clockwise solitonic vortex (3)
    \item[E:] counterclockwise solitonic vortex (4) 
    \item[F:] canted (5) \vspace{-0.5\baselineskip}\mbox{}
\end{itemize} & List\\
\texttt{excitation\_quality} & Quality metric for excitations in the lone excitation class  & List \\
\texttt{soldet\_CNN} & CNN classifier label from SolDet & Integer \\
\texttt{soldet\_OD} & List of positions returned by SolDet & List \\
\texttt{soldet\_PIE} & List of classes returned by PIE classifier for all excitation localized in \texttt{soldet\_OD} & List \\
\texttt{soldet\_QE} & List of quality estimates for all excitation localized in \texttt{soldet\_OD} & List \\
\bottomrule
\end{tabular*}
\label{tab:data-dictionary}
\end{table}

In the next stage of data curating, we further refine labels for the data that are not in class-8 or class-9 in \texttt{label\_v2} using five distinct deep ensembles of size ten trained through a repeated five-fold cross-validation.
Prior research suggest that ensembles of ten models are sufficient to reliably assess the predictive uncertainty~\cite{Lakshminarayanan17-UDE}. 
Building on that, we use ten five-fold-cross-validated OD models\footnote{There were 50 models in total---ten per cross-validation---each trained on $80~\%$ of the data. 
To ensure diversity of models, the dataset was shuffled between the ten training sessions.} trained using \texttt{label\_v2} class-0 and class-1 ($4\,683$ images in total).

Each image is tagged with ten OD predicted labels, each consisting of the number of excitations detected and their positions.
Given the random initialization of the training sessions, we treat the deep ensemble as giving a uniformly-weighted set of predicted labels, with each model prediction considered equally reliable~\cite{Lakshminarayanan17-UDE}.
The OD class predictions are used to define a measure of class-based disagreement
\begin{equation}
    D_{class} = \frac{\#(\text{OD class prediction} \neq \texttt{label\_v2})}{M},
\end{equation}
where $\#(\cdot)$ denotes the number of instances for which condition $(\cdot)$ occurs and $M$ is the size of the deep ensemble.
A score of $0$ indicates full agreement within the deep ensemble with the ground truth label while score of $1$ indicates that all models predict an incorrect class. 
In addition, each image in class-1 is assigned a preliminary excitation position (${\rm FIT}_{\rm pos}$): the minimum of the background subtracted 1D density profile~\cite{Fritsch20-SCV,Guo21-MDS}.

The subsequent data analysis continues our aim of identifying mislabeled images and is carried out separately for class-0, 1, and 2.
The result of this analysis is stored in \texttt{label\_v3}.

\begin{description}
    \item[{\rm Class-0:} {\it Deep ensemble disagreement.}] 
    For each image, we calculate the class-based disagreement score $D_{\rm class}$.
    Since the initial agreement rate between annotators was around $87~\%$, we opt to require at least $90~\%$ agreement between the ODs. 
    Thus, images with $D_{\rm class-0}\leq0.1$ are retained in the curated dataset with \texttt{label\_v3} set equal to \texttt{label\_v2} ($1\,130$ images).
    We note that the remaining $104$ images had either (a) $D_{\rm class}\in(0.1, 0.5]$ ($75$ images); or (b) half or more models predicting class-1 ($29$ images).
    All of these are assigned to class-8.
    \item[{\rm Class-1:}]
    \begin{description}
        \item[{\it Step 1: Deep ensemble disagreement.}]
        Like for class-0, we first compare the number of OD-identified excitations with \texttt{label\_v2} and find $3\,222$ images for which $D_{\rm class}\leq0.1$.
        Of the remaining $227$ there are $178$ for which $D_{\rm class}\in(0.1, 0.5]$, $28$ images for which five or more models predict class-0, and $21$ images in which five or more models assign class-2.
        All these images are assigned to class-8.
        \item[{\it Step 2: Position range check.}]
        While the ODs agree on the number of solitonic excitations, the positions in pixel units found by the ODs might differ. 
        All images for which $\max(\textbf{OD}_{\rm pos})-\min(\textbf{OD}_{\rm pos})>3$ are assigned to class-8 ($10$ images)\footnote{This threshold corresponds to the average width of solitons in our images which we observed to be about four pixels.}.  
        Here $\textbf{OD}_{\rm pos}$ is the vector of OD position predictions and $\max(\textbf{v})$ and $\min(\textbf{v})$ denote the maximum and minimum element of the of the vector $\textbf{v}$, respectively.
        \item[{\it Step 3: Position alignment check.}]
        As an additional consistency check, we compare the average OD position for images against the preliminary ${\rm FIT}_{\rm pos}$ position. 
        Images for which $|\overline{\textbf{OD}}_{\rm pos} - {\rm FIT}_{\rm pos}|>3$ ($15$ images) were manually reviewed.  
        We found that in each case the excitation position was better located by the OD, so the excitation position label was updated to that returned by the OD.
    \end{description}
    \item[{\rm Class-2:}]
    \begin{description}
        \item[{\it Step 1: Deep ensemble disagreement.}]
        Data in class-2, by design, includes ambiguous images.
        Thus, the goal of curating this class is to ensure that it does not contain images that would with high confidence be classified as belonging to either class-0 or class-1 by the deep ensemble.
        Since all models were trained only on the class-0 and class-1 data, we use an ensemble consisting of all 50 models.
        We find that out of the $1\,388$ images in class-2, $30$ had 90~\% of models predict class-0 (and were therefore assigned to class-8) and $336$ had 90~\% of models predict class-1 (and were further analyzed in step 2).
        The remaining $1\,022$ images are retained in the curated dataset with \texttt{label\_v3} set equal to \texttt{label\_v2}.
        \item[{\it Step 2: Position range check.}]
        To confirm the OD predicted class-1, we compare the range of OD positions for those models predicting one excitation and find that $\max(\textbf{OD}_{pos})-\min(\textbf{OD}_{pos})<3$ for $322$ of images, suggesting that these are very likely class-1 data mislabeled as class-2.
        These images are also assigned to class-8.
    \end{description}
\end{description}

The deep-ensembles-based data curating process resulted in assigning to class-8 a total of $693$ images from the original dataset.
The resulting dataset contains $1\,130$ images in class-0, $3\,212$ in class-1, $1\,036$ in class-2, and $879$ in class-8 (data labeled as class-9, associated with unlabeled data, are unchanged).
The final classifications are contained in \texttt{label\_v3.}

\subsection{Label refinement: dark solitons in BECs dataset v.2.0}\label{sec:labeling-pie}

To further refine labels for images in class-1, we use the PIE classifier and quality estimator from the SolDet package~\cite{SolDet}.
PIE classifier partitions class-1 into physically-motivated sub-classes stored in \texttt{excitation\_PIE}.
The PIE classifier operates by splitting each image into top and bottom halves and determining the associated 1D profiles to which the quality estimator is separately applied.
In addition to returning an overall quality estimate, stored in \texttt{excitation\_quality}, this algorithm also returns parameters such as the excitation position, width, and so forth.

Then, a sequence of thresholds driven by different top-bottom combinations of these parameters is applied to determine the label.
The values defining all thresholds were arrived at by exploring the data accepted and rejected by the cut to minimize the false positive identification of longitudinal solitons, as described in reference~\cite{Guo22-CMP}.

Within the $3\,212$ images in class-1, the PIE classifier categorized $2\,229$ images as proper longitudinal solitons (class-A).
Out of the remaining images, $378$ were classified as top ``partial'' solitons (class-B) and $418$ as bottom ``partial'' solitons (class-C); $28$ were categorized as clockwise vortex (class-D) and $38$ as counterclockwise vortex (class-E); $121$ were categorized as canted excitations (class-F).

\section{Automated expansion of the dataset v.1.0}

In this section, we describe how we leverage the full SolDet package~\cite{SolDet} to automatically analyze and label previously unlabeled images (a subset of about $1\times10^4$ identified in \texttt{label\_v3} as class-9) that were not part of {\it Dark solitons in BECs dataset v.1.0}.
These data include images representing class-0, 1, and 2; many class-2 candidates possess multiple excitations as well as ambiguous and confusing structures that may hinder human labeling.
Since these data were not previously considered, they make an ideal test case for the SolDet package~\cite{Guo22-CMP}.
Roughly $90~\%$ of the class-0, 1, and 2 data from the curated dataset were used to train SolDet, leaving the remaining $10~\%$ of these classes for validation.
In addition, we also apply SolDet to all of class-8 allowing us to cross-check the mislabeled assignment.

\begin{figure}[t]
    \centering
    \includegraphics[width=1.0\linewidth]{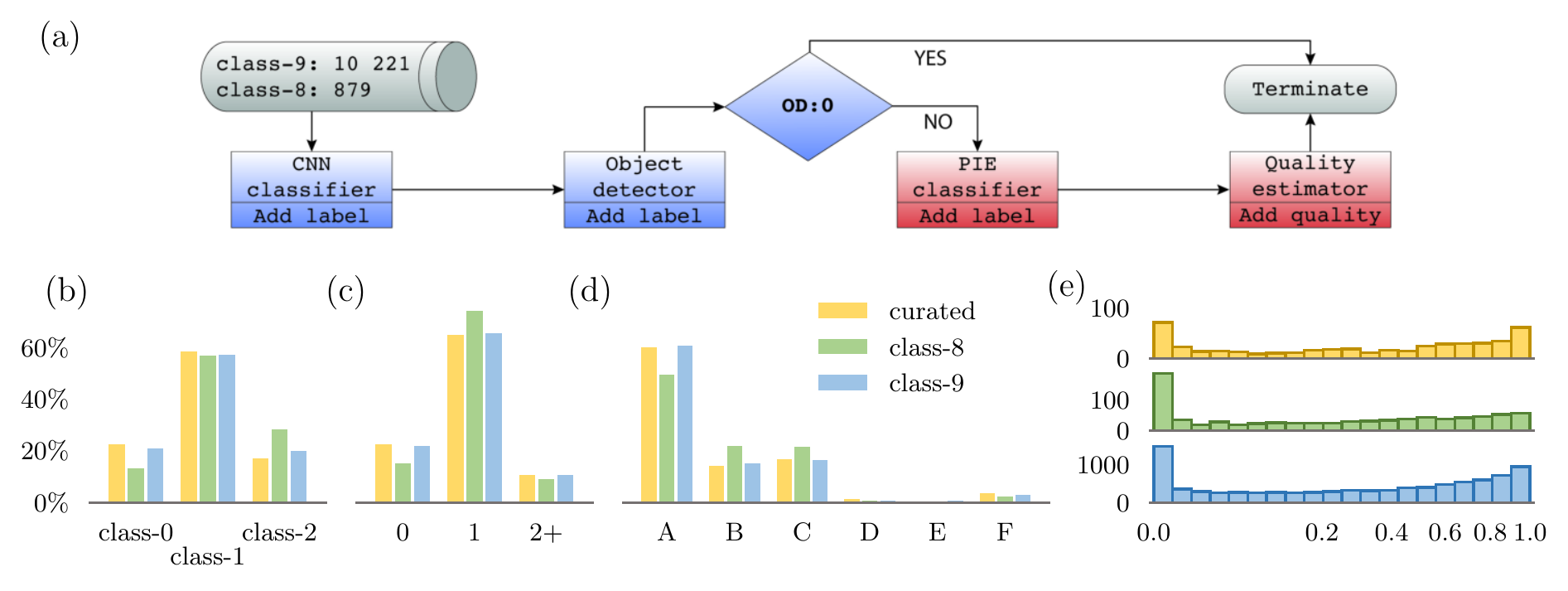}
    \caption{SolDet flow chart labeling results.
    (a) A high-level view of the SolDet algorithm (adapted from reference~\cite{Guo22-CMP}), where black arrows follow the data flow between blocks describing each analysis stage as well as added labels.
    The bottom four panels histogram the results from each stage of the SolDet labeling process: (b) CNN classification; (c) number of excitations identified by OD; (d) type of excitation identified by the PIE classifier; and (e) quality metric of all detected excitations.
    These panels show the 10~\% curated data set aside for validation (yellow), the mislabeled class-8 data (green), and the previously unlabeled class-9 data (blue). 
    }
    \label{fig:soldet-flow}
\end{figure}

For this application, the SolDet package is configured to give the labeling flow depicted in figure~\ref{fig:soldet-flow}(a), involving a sequential application of CNN and OD modules followed by the PIE classifier.
The CNN module categorizes images as class-0, 1, or 2 while the OD module tags each image with positions of all detected excitations.
If OD assigns class-0, the image is labeled accordingly and the process terminates for that image. 
Otherwise, the PIE classifier is executed for each excitation detected by the OD.
Finally, all excitations located by OD are additionally tagged with a quality estimate~\cite{Guo22-CMP}.

The output of each module is included as a separate label: \texttt{soldet\_CNN} for the CNN classifier; \texttt{soldet\_OD} for the vector of positions; \texttt{soldet\_PIE} for the vector of classes returned by PIE classifier; and \texttt{soldet\_QE} for the vector of quality estimates.
This enables the end user to choose a desired level of agreement between the labeling modules or the longitudinal solitons' quality necessary for a particular application. 
Thus, unlike images in the curated dataset, these previously unlabeled images are not assigned a single ground truth class. 

Figures~\ref{fig:soldet-flow}(b)--(d) compares the label assignment for three different subsets of our dataset: a subset of the curated data ($10~\%$) automatically generated by SolDet for validation (shown in yellow), class-8 (mislabeled data; shown in green), and class-9 (unlabeled data; shown in blue).
Panel (b) shows that CNN classes for each case follow a very similar distribution, but where data in class-8 has a reduced likelihood of being classified in class-0.
While the OD results depicted in panel (c) are very similar to the data in (b), there are slight differences.
For example in the class-8 data, the OD finds somewhat more class-1 excitations than the the CNN did (by about $17.3~\%$). 
This likely results from the annotators disagreeing on the labeling of partial solitons and solitonic vortices, all of which are identified as a single excitation by the OD.
The PIE classifier in (d) finds a modest deficit of class-A (longitudinal soliton) in the mislabeled data, confirming that such excitations are the most straightforward to annotate.
Lastly, figure~\ref{fig:soldet-flow}(e) shows that the quality metric for all excitations identified by the OD are consistent across the three subsets of our data.

\begin{figure}[t]
    \centering
    \includegraphics[width=1.0\linewidth]{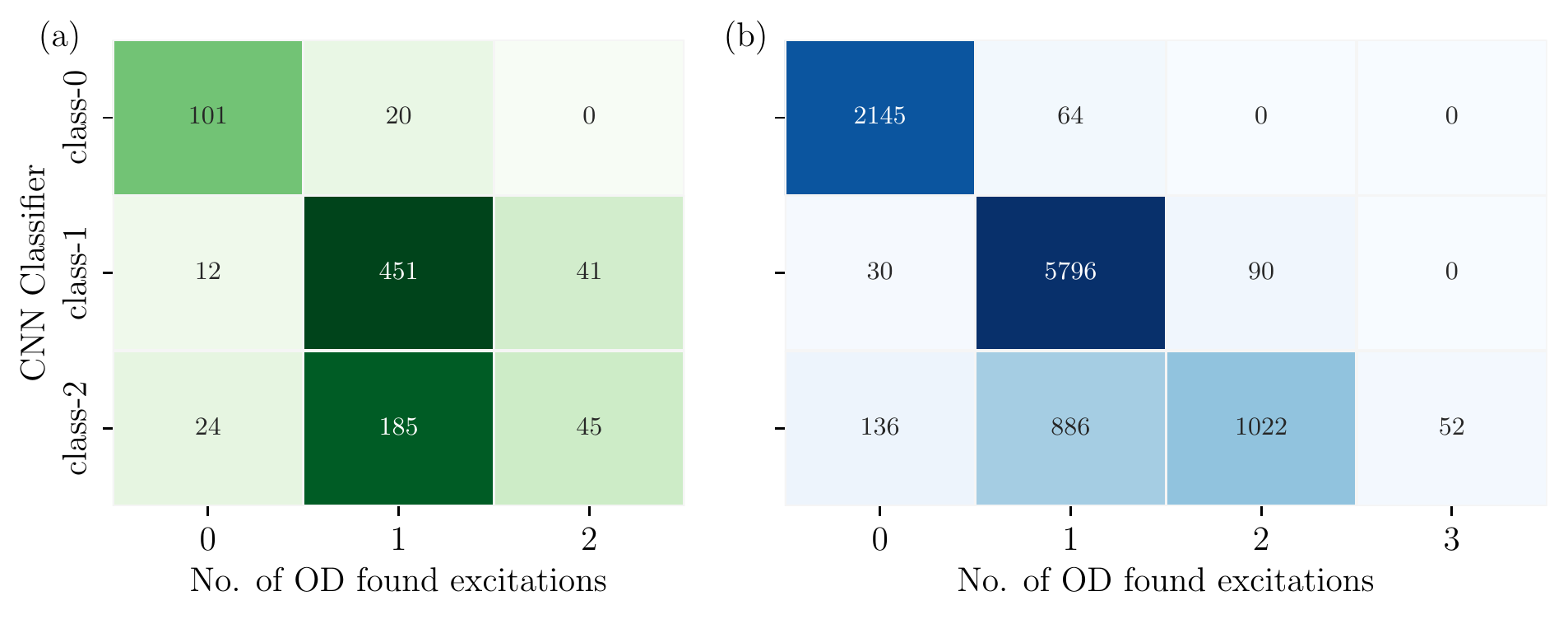}
    \caption{Matrices comparing the outputs of the CNN module and the OD module for (a) the mislabeled class-8 data as well as (b) the previously unlabeled class-9 data. 
}
    \label{fig:data-stat}
\end{figure}

Figure~\ref{fig:data-stat} compares the outcome of the CNN and OD modules for (a) class-8 data and (b) class-9 data.
In both cases when the CNN classifier assigns class-0 or class-1, the OD is unlikely to find any or other number than one excitation, respectively.
However, when the CNN assigns class-2, the OD assignment is strongly biased to class-1 for the class-8 (mislabeled) data, with $73~\%$ assessed to contain only one excitation. 
This bias likely results from the process of labeling and curation in which many class-1 candidates were moved into class-8 to avoid false positives. 
The OD assignment seems almost random for the previously unlabeled class-9 data assigned class-2 by the CNN, with $42~\%$ assessed to contain only one excitation and $49~\%$ assessed to contain two excitations.

For the mislabeled (class-8) data the performance is significantly degraded in the converse case: about $30~\%$ of the OD class-0 and 1 data and almost $50~\%$ of OD class-2 data is assigned one of the alternative CNN classes.
For the unlabeled (class-9) data, the disagreement between OD and CNN assigned classes is much lower, at $7~\%$, $14~\%$, and $8~\%$ for OD class-0, 1, and 2+, respectively.

\begin{figure}[t]
    \centering
    \includegraphics[width=\linewidth]{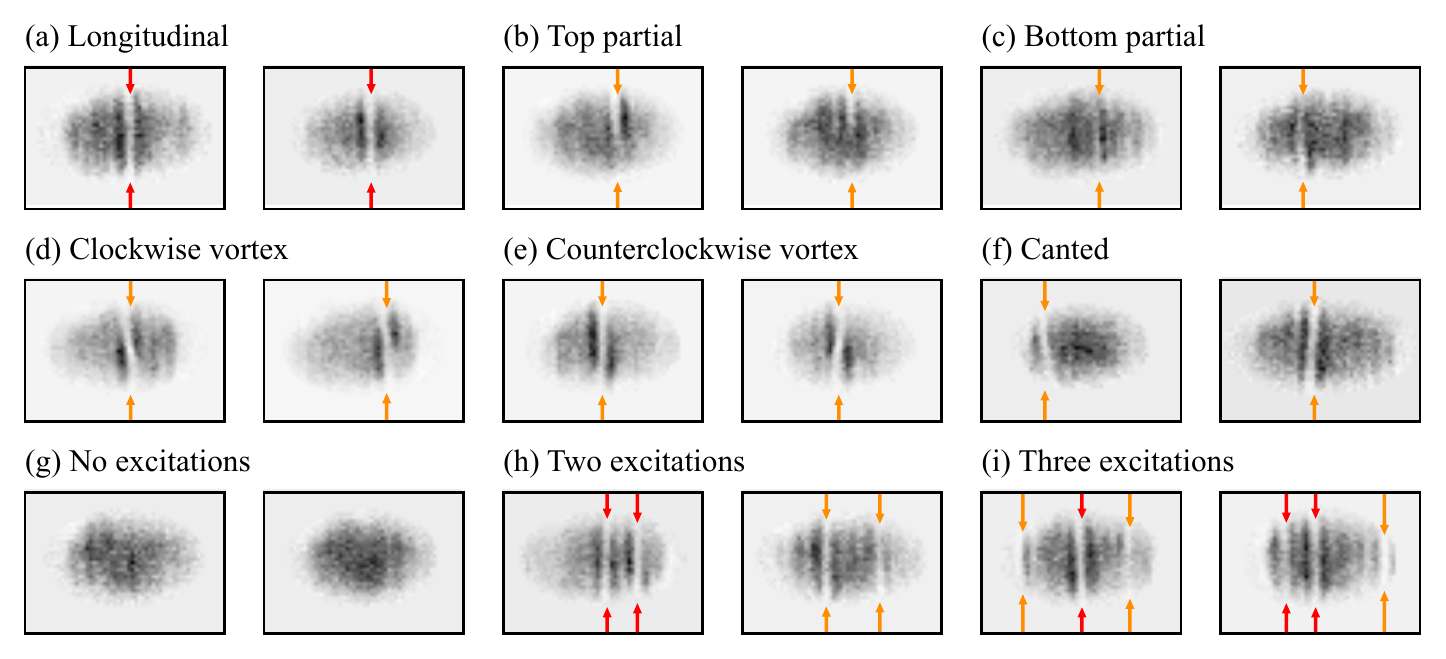}
    \caption{Representative class-9 data from the {\it Dark solitons in BECs dataset v.2.0}.
    (a) longitudinal soliton; (b) top partial; (c) bottom partial;
    (d) clockwise solitonic vortex; (e) counterclockwise solitonic vortex; (f) canted excitation; (g) no excitation; (h) two excitations; and (i) three or more excitations.
    Red arrows indicate longitudinal solitons while orange ones are for other solitonic excitations.
}
    \label{fig:sample-unlabel-data}
\end{figure}

Figure~\ref{fig:sample-unlabel-data} depicts 18 images, with a variety of classifications and positions, all from the automatically labeled (class-9) portion of the expanded dataset. 
By design, SolDet effectively categorizes class-1-A images, giving the examples in (a).
Panels (b)--(f) show illustrative examples from the remaining classes.
Because the primary function of the PIE classifier is to reject images that are not class-1-A (avoid false positive longitudinal solitons) the B--F labels are of lower quality.
The bottom row displays data from class-0 (no excitations; panel (g)) and class-2 (other excitations; panels (h)--(i)). 
Together, these show that SolDet is very effective in delineating between class-0,1 and 2.

In each panel the arrows identify the location of the excitation from the OD, showing it is effective across-the-board in locating excitations.  
The arrows are colored according to the PIE classifier result: Red arrows mark the location of longitudinal solitons (class-*-A) and the orange arrows mark all other classes.
Even in cases with many excitations (h), (i), SolDet correctly identifies high quality longitudinal solitons.

\section{Conclusion and outlook}\label{sec:conclusion}

We find that SolDet reliably labels new experimental data (class-9) as well as data categorized as potentially mislabeled (class-8).
An inspection of the images shows that the the assigned classes are visually similar to the manually labeled data.
Furthermore, the labels automatically assigned to class-9 have a statistical distribution that is very similar to the training dataset.
By contrast, the labels assigned to class-8---a selectively filtered subset of the original dataset---are significantly different.

As we mentioned, SolDet is configured to identify and correctly locate longitudinal soliton within BEC (class-*-A).
The reliability of the additional PIE classes could be improved by, e.g., further refining cuts defining the physically-motivated categories or slicing the image into more than two pieces.

The enlarged {\it Dark solitons in BECs dataset v.2.0} dataset includes quantitative estimates of all longitudinal solitons quality as well as new fine-grained solitonic excitation categories of all detected excitations.
It is a freely available to the whole ML and physics community the opportunity to develop novel ML techniques to cold atom systems and to further explore the intersection of ML and quantum physics.

\section*{Data availability statement}
The data that support the findings of this study are openly available at the following URL/DOI: \href{https://doi.org/10.18434/mds2-2363}{https://doi.org/10.18434/mds2-2363}.

\section*{Acknowledgments}\label{sec:Acknowledgment}
This work was partially supported by the National Institute of Standards and Technology and the National Science Foundation through the Physics Frontier Center at the Joint Quantum Institute and the Quantum Leap Challenge Institute for Robust Quantum Simulation.
The views and conclusions contained in this paper are those of the authors and should not be interpreted as representing the official policies, either expressed or implied, of the U.S. Government. 
The U.S. Government is authorized to reproduce and distribute reprints for Government purposes notwithstanding any copyright noted herein. 

\renewcommand{\thesection}{A}
\renewcommand{\thefigure}{A\arabic{figure}}
\setcounter{figure}{0}
\renewcommand{\thetable}{A\arabic{table}}
\setcounter{table}{0}
\renewcommand{\arraystretch}{1}
\section{Appendix}\label{sec:appendix}
\subsection{Experimental Setup}\label{app:exp-setup}
In our experiments, we implement well established techniques to achieve the Bose--Einstein condensation of neutral atoms \cite{ketterle1999making}. 
Our $^{87}\text{Rb}$ BECs contain $N\approx2.4\times 10^5$ atoms in an elongated crossed optical dipole trap with harmonic frequencies $\left[ \omega_x,\omega_y,\omega_z\right]= 2\pi\times\left[9.1(1),94.5(6),153(1)\right]$ Hz.

To create solitons, we manipulate the BEC local density and phase by applying an external potential generated by a far detunned laser light patterned by a digital micromiror device (DMD). 
The DMD (Texas Instruments DLP LightCrafter Module---DLP3000)\footnote{Certain commercial equipment, instruments, or materials are identified in this paper in order to specify the experimental procedure adequately. 
Such identification is not intended to imply recommendation or endorsement by the National Institute of Standards and Technology, nor is it intended to imply that the materials or equipment identified are necessarily the best available for the purpose.} has an array of $608\times684$ mirrors, $\approx 7.6\ \mu\rm m$ on a side, that can be independently flipped to create arbitrary patterns.
The pattern generated by the DMD is imaged onto the atoms using an imaging system that demagnifies the light $12$ times and the intensity of the laser light is controlled by an acousto-optic modulator. 

The protocol to create solitons, which starts after the BEC is formed, is summarized as follows: with the DMD previously programmed to reflect a narrow 3 px wide stripe, we increase the laser power creating a dimple potential that depletes the BEC local density by about $70~\%$. 
We then change the DMD pattern to illuminate half of the BEC extension and pulse the light for a variable time to imprint the phase. 
Since the accumulated phase is proportional to the duration in which the light is pulsed, the pulse duration is varied to create solitons at different speeds. 
To avoid creating additional density modulations after imprinting the phase, the DMD is reconfigured back to the narrow stripe, the dimple is reapplied and its magnitude is ramped to zero. 
More details about this protocol can be found in \cite{Fritsch20-SCV}.

After solitons are created we let them oscillate in the trap for a variable evolution time that allow us to obtain variation in the soliton properties, such as the oscillation amplitude, initial position, propagation velocity, and lifetime. 
Since our elongated trap geometry does not produce truly 1D BECs, kink solitons that are initially created can eventually decay into solitonic vortices during the time they oscillate in the trap. 
After the evolution time, the trap is turned off and the cloud expands for $12\ {\rm ms}$ before it is imaged using standard absorption imaging \cite{ketterle1999making}. 

In the standard absorption imaging technique, the fraction of a resonant probe light absorbed by the atomic cloud is used to extract information about the atoms. 
For the standard absorption imaging, we acquire three images that are combined to obtain the optical density. 
In the first image $I^{\rm A}_{i,j}$, the atomic cloud is illuminated with the probe light and its shadow is recorded in a CCD camera, see figure~\ref{fig:soliton-sample}(a). 
To compute the fraction of the light that is absorbed by the atoms, a second image $I^{\rm P}_{i,j}$ of the probe beam, without atoms, is acquired, figure~\ref{fig:soliton-sample}(b). 
The third image $I^{\rm BG}_{i,j}$ is then acquired without the probe light to get the background light in the experiment, figure~\ref{fig:soliton-sample}(c). 
All images are acquired with the same duration and the probe beam has the same intensity for the first two images. 
The three images are then combined to to obtain the 2D optical density
\begin{equation}\label{eq:od}
    \sigma_0 n_{i,j} \approx -\ln\left[\frac{I^{\rm A}_{i,j}-I^{\rm BG}_{i,j}}{I^{\rm P}_{i,j}-I^{\rm BG}_{i,j}}\right],
\end{equation}
where $\sigma_0=3\lambda^2 / (2\pi)$ is  the resonant cross-section and $\lambda$ is wavelength of the probe laser. 

All images [figures~\ref{fig:soliton-sample}(a)--(c)] are acquired by a $648\times488$ pixel camera (Point Grey FL3) with $5.6\ \si{\micro\meter}$ square pixels, labeled by $i$ and $j$. 
The imaging system, with an optical resolution of $\approx 2.8\ \si{\micro\meter}$, has $\approx 6\times$ magnification, generating images with effective pixel size of $0.93\ \si{\micro\meter}$. 

\bibliographystyle{unsrt}
\newcommand{\newblock}{}

\end{document}